**Ionic Interdiffusion at Cathode|Solid-Electrolyte Interface:** A Machine Learning–Assisted Multiscale Investigation and Mitigation Strategies

Musawenkosi K. Ncube[a], Pallab Barai[b], Selva Chandrasekaran Selvaraj[a], Larry A. Curtiss[b], Anh T. Ngo[a,b,*], Venkat Srinivasan[b,*]

[a]Department of Chemical Engineering, University of Illinois Chicago, Chicago, IL 60608, United States
[b]Materials Science Division, Argonne National Laboratory, Lemont, IL 60439, United States

Corresponding authors: **anhngo@uic.edu ; vsrinivasan@anl.gov**

**Abstract**

Future lithium-based batteries are expected to use solid electrolytes to achieve higher energy density and fast charge capabilities. The majority of solid electrolytes are thermodynamically unstable against layered oxide cathodes. Here, the stability of $LiCoO_2$ (LCO) cathode with $Li_{10}GeP_2S_{12}$ (LGPS) solid electrolyte is investigated using *ab* initio molecular dynamics (AIMD) and machine learning molecular dynamics (MLMD). The propensity of ionic interdiffusion, formation of a passivation layer, and corresponding decay in cell performance is addressed using a continuum model. The large-scale MLMD simulations confirm that the LCO|LGPS interface permits interdiffusion of Co and other ionic species, leading to the formation and growth of a resistive interphase and dramatic capacity fade even in the first cycle. We then examine the literature evidence that incorporating a thin layer of $LiNb_{0.5}Ta_{0.5}O_3$ (LNTO) between LCO and LGPS prevents the interdiffusion of ions. Atomistic simulations suggest that the substitution of Li in LNTO with Co is not thermodynamically favorable, which helps to minimize the ionic interdiffusion process. The stable Nb/Ta$^{5+}$ states form a rigid metal-oxide framework, which consequently also prevents the substitution of Nb/Ta. However, continuum level analysis suggests that due to the higher mechanical stiffness of LNTO, interfacial delamination between the LCO and LNTO is possible, which can minimize the effectiveness of the protective layer. This paper suggests the need for the development of novel interlayers that balance low interdiffusion with low stiffness.

**Introduction**

All-solid-state lithium batteries (ASSLBs) are promising next-generation energy storage systems[1,2]. They replace the flammable organic liquid electrolytes used in commercial batteries with inorganic solid-state electrolytes (SSEs). The growing interest in ASSLBs stems from the limitations of commercial lithium-ion batteries (LIBs), which are approaching the theoretical specific capacity of 372 mAh/g observed in the graphite anodes[3,4]. Additionally, LIBs have safety concerns, high-capacity fading, and issues related to dendrite formation[5–8]. The development of ASSLBs aims to address these limitations while also offering higher energy densities. The integration of inorganic SSEs in ASSLBs also improves ionic conductivity, safety, and cycle life[9,10]. Among the SSEs, sulfide-based electrolytes have garnered significant attention for their high ionic conductivities (ranging from $10^{-2}$ to $10^{-5}$ S/cm at room temperature), which are comparable to those of organic liquid electrolytes (LE)[11–13]. They also exhibit good mechanical strength, flexibility, and low grain boundary resistance, making them ideal candidates for ASSLBs[7,12,14]. One of the most studied inorganic SSEs is $Li_{10}GeP_2S_{12}$ (LGPS), which has a high $Li^+$ ionic conductivity (12 mS/cm) at room temperature[15,16]. LGPS also possesses favorable mechanical properties, making it a preferable material over oxide-based SSEs[10,14,16,17]. However, there are interfacial issues observed when LGPS is paired with high-voltage cathode materials like $LiCoO_2$ (LCO), particularly cobalt (Co) diffusion to the LGPS electrolyte, which detrimentally affects the battery performance[18–21]. To mitigate this challenge, researchers have explored the introduction of thin oxide interphase layers between sulfide-based SSEs and LCO[7,12,18]. This approach prevents direct contact between LGPS and LCO, minimizing mutual ion diffusion and interfacial resistance.

Li-Nb-Ta oxides ($LiNbO_3$ (LNO) and $LiTaO_3$ (LTO)) are some of the commonly investigated interphase materials due to their high ionic conductivity, and stability against sulfide-based electrolytes[18,22–25]. Further studies done on the Li-Nb-Ta ternary oxide system also revealed the potential of a new material, $LiNb_{0.5}Ta_{0.5}O_3$ (LNTO), to deliver higher $Li^+$ mobility and permittivity compared to LNO and LTO[26]. Zhang, W. *et al.,* demonstrated promising results using LNTO in reducing interfacial resistance and improving the electrochemical performance of ASSLBs[7,27,28]. However, the experimental findings do not detail the chemistry behind the improved performance displayed by LNTO in mitigating the cobalt interdiffusion problem. In their other paper, Zhang et al also observe Co diffusion after long-term electrochemical cycling, which means LNTO does not fully suppress the phenomena[27]. Our work, therefore, takes a computational approach to extend the experimental studies on LNTO, with an aim to explain the chemistry of the interface using multiscale simulations. We carry out molecular dynamics and continuum-level calculations to understand the behavior of LGPS and LNTO electrolyte materials and their interface properties with LCO. Due to the atomic-level complex interplay between physicochemical and electrochemical properties at the interface, our study requires large-scale atomic simulations to adequately investigate Co interdiffusion. AIMD simulations, which are known for accurately predicting interfacial electrochemical properties, are computationally intensive and limited to a few hundred atoms, and therefore cannot be efficiently used for extended simulation times ≥ nanosecond scale [8,29–33]. In this paper, we integrate AIMD with Machine Learning Molecular Dynamics (MLMD) to overcome this challenge, and we run our calculations using systems with greater than a thousand atoms for elongated times[31]. The Deep Learning Potential (DLP) based on the neural network method is adopted for this study as it is an efficient machine learning technique that maintains the accuracy of AIMD while boosting the time and size scale of the system[29,34,35]. We use the diffusion coefficients obtained from atomistic simulations to develop a continuum-level model that allows the investigation of two types of interfacial degradation mechanisms: (i) Delamination, and (ii) Interdiffusion induced passivation[36].

By integrating AIMD, MLMD, and the continuum model, we aim to provide a comprehensive understanding of the interfacial phenomena, particularly focusing on Co migration at LCO|LGPS and LCO|LNTO interfaces, and develop a multiscale theoretical approach for studying ion diffusion and stability in ASSLBs. We investigate the thermodynamic and physical stability of the two solid-solid interfaces to explain the experimental observations discussed above. This work is organized in the manuscript as follows: we start with a description of the simulation methods used, which include AIMD, MLMD, and the continuum model. The results and discussion section then describes the diffusion coefficients of species in LCO|LGPS and cobalt interdiffusion at the LCO|LGPS and LCO|LNTO interfaces, as well as the delamination behavior in the two interfaces. Finally, we conclude by outlining the key findings, recommendations, and future works.

**Computational Methodology**

**DFT calculations**
Spin-polarized DFT calculations were employed to optimize the initial bulk structures and surfaces for LCO, LNTO, and LGPS. The Projector-Augmented Wave (PAW) potentials for the elements were implemented as supplied by the Vienna *Ab Initio* Simulation Package (VASP)[37,38]. The Generalized Gradient Approximation(GGA), based on the Perdew–Burke–Ernzerhof approach, accounted for the exchange-correlation energies[39]. A DFT+U approach described by Dudarev et al. is used for LCO, with a Hubbard U correction for Co set to 5.9 eV based on literature studies, accounting for the correlation effect of Co-3d electrons[40,41]. The kinetic energy cutoff was set to 600 eV to enhance accuracy, and the Brillouin zone was sampled at the gamma point with the *k*-mesh set based on the respective lattice parameters. The energy and force convergence thresholds were set to less than ±10 meV/Å.

The bulk structures for $LiCoO_2$ (LCO), $Li_{10}GeP_2S_{12}$ (LGPS), and $LiNbO_3$ (LNO) were sourced from the Materials Project[42] and minimized to the ground state energy. The surfaces 100, 104, and 110 for LCO [13,27,43], and 100 for LGPS [44] chosen based on the stable structures in the cited studies, were cut out from replicas of the unit cells, using VESTA[45]. LNO was doped with Tantalum (Ta) to the ratio $LiNb_{0.5}Ta_{0.5}O_3$. Figure 1(a-c) illustrates the LCO, LGPS, and LNTO bulk structures. To determine the most stable LNTO structure, the energies of 18 surfaces were calculated using equation 1[46]. The lowest energy surface (131), as determined by comparing the surface energies in Figure 1d, was used in this study. VESTA and OVITO were used in this work to visualize the simulation structures and trajectories[45,47].

$$\sigma = \frac{1}{2A}(E_{slab} - n_{slab}E_{bulk}) \text{ --------------- (1)}$$

Where; *A* is the surface area; $E_{slab}$ is the energy of the slab; $n_{slab}$ is the number of formula units in the surface slab; $E_{bulk}$ is the energy of one unit of the bulk structure.

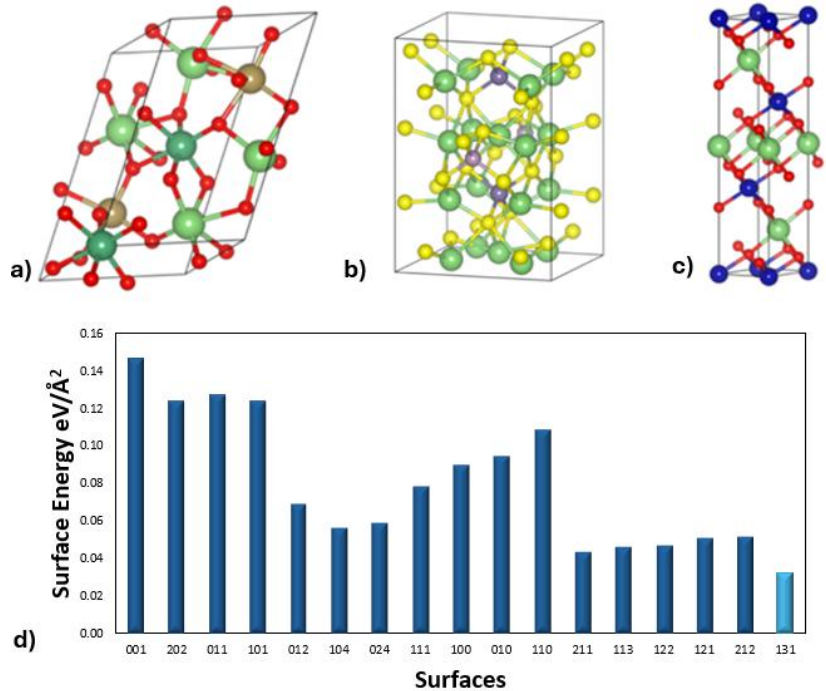

Figure 1: Ball and stick models of optimized bulk structures of a) LNTO, b)LGPS, and c) LCO; Red, blue, light green, dark green, tan, yellow, purple, and gray balls are O, Co, Li, Nb, Ta, S, Ge, and P atoms, respectively; d) Surface Energies for 18 surfaces cut out from the $LiNb_{0.5}Ta_{0.5}O_3$ bulk structure

The interfaces of less than 300 atoms for LCO|LPGS and LCO|LNTO were built with the Virtual NanoLab (VNL) software in the QuantumATK package[48] and subjected to AIMD simulations. The calculations were conducted within a canonical NVT ensemble, and a Nosé–Hoover thermostat was used to keep the temperature of the system constant[49]. The AIMD simulations were performed for at least 5 picoseconds (ps) with a time step of 1 femtoseconds (fs). Variations in temperature(from 300K to 900K), cell size (from -20% to +20% of cell volume), and the effect of vacuum were considered to generate sufficient data to train a neural network-based, accurate machine-learning model.

**Deep Learning Potential (DLP) ;**

The DLP for the machine learning molecular dynamics simulation was developed using a Deep Neural Network (DNN), an artificial multi-layer neural network. The algorithm for the DNN method was implemented in DeePMD-kit (v2.2.7)[50], which was designed using the TensorFlow framework. The smoothing version of the deep potential, an end-to-end machine-learning-based interatomic Potential Energy Surface (PES) model[51] was used, with a cutoff radius for adjacent atoms set to 6 Å. It has been proven to efficiently represent the PES of various systems with the accuracy of DFT calculation[29].

To develop the DLP, the total interatomic energy of the system, $E$, is taken as a function of atomic coordinates, represented by the local coordination matrix $R$ (equation 2), and $E$ is the sum of atomic energy contributions (equation 3)[29,51–53].

$$R = \{r_1^T, \ldots, r_i^T, \ldots, r_N^T\}^T \quad \text{---------------- (2)}$$

Where ; $r_i = (x_i, y_i, z_i)$ in a 3D Euclidean space and $N$ is the total number of atoms.

$$E = \sum_i E_i \quad \text{---------------- (3)}$$

$E_i$ is computed using a local atomic environment matrix defined by $R_{ij}$ ;

$$\{R_{ij}\}_{i=1}^{N} = \{r_{i1}^T, \ldots, r_{i2}^T, \ldots, r_{ij}^T | j \in N_{Rc}(i, R_c = 6.0\}^T \quad \text{---------------- (4)}$$

Where ; $N_{Rc}(i)$ is the index set of the $i^{th}$ atom's neighbors within the cutoff radius $R_c$ i.e., the relative coordinate $r_{ij} = r_i - r_j \leq R_c$

An embedding neural network with three layers, each containing 25, 50, and 100 neurons, was used to generate a symmetry-preserving descriptor $D_i = D_i\left(\{R_{ij}\}_{i=1}^{N}\right)$ based on the local atomic coordinates $R \in \mathbb{R}^{N \times 3}$ [29,50]. The descriptor values were then passed to the fitting neural network with three layers, each comprising 240 neurons. This fully connected feedforward DNN then outputs $E_i$, used by the model to calculate total energy and force as described by Selva et al[29]. This workflow is simplified by an illustration in Figure 2. The model was trained for $3 \times 10^5$ steps using the data set from the AIMD and DFT runs. The loss function was determined at every step as a function of the mean square error for energy and forces[54]. These are tabulated in Table 1 for the LCO|LGPS and LCO|LNTO interfaces. The learning rate was set to decrease exponentially relative to the starting value of $10^{-3}$ and stopping at $3.51 \times 10^{-8}$, with a decay step of 5000. The results of the predicted forces and energy data vs the trained data are plotted in Figure 3 (LCO | LGPS) and Figure S1 (LCO | LNTO).

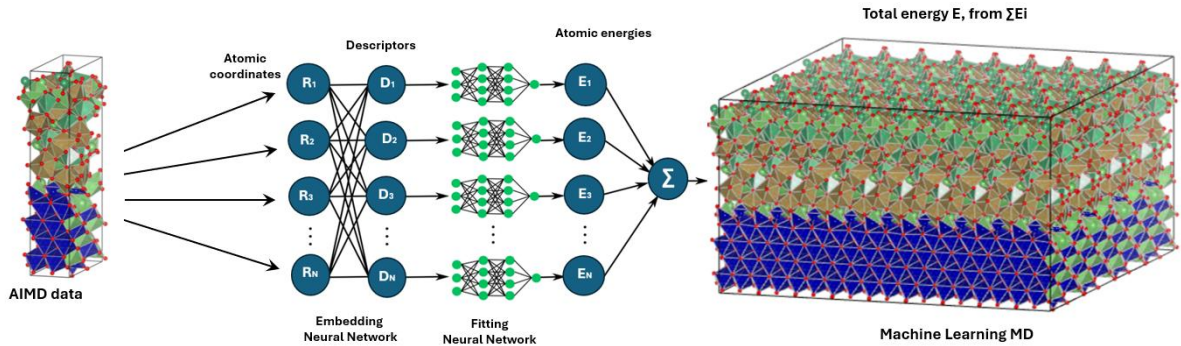

Figure 2: Schematic of the workflow in developing a Deep Learning Potential for Machine Learning Molecular Dynamics.

Table 1: The calculated loss function parameters as a function of the Mean Absolute Error(MAE) and Root Mean Square Error(RMSE) of the DLP model

|  | Unit | LCO|LGPS | LCO|LNTO |
| --- | --- | --- | --- |
| Energy MAE | eV/atom | $4.8 \times 10^{-5}$ | $1.57 \times 10^{-3}$ |
| Energy RMSE | eV/atom | $6.4 \times 10^{-5}$ | $3.58 \times 10^{-3}$ |
| Force MAE | eV/Å | $1.8 \times 10^{-3}$ | $1.45 \times 10^{-1}$ |
| Force RMSE | eV/Å | $2.8 \times 10^{-3}$ | $2.02 \times 10^{-1}$ |

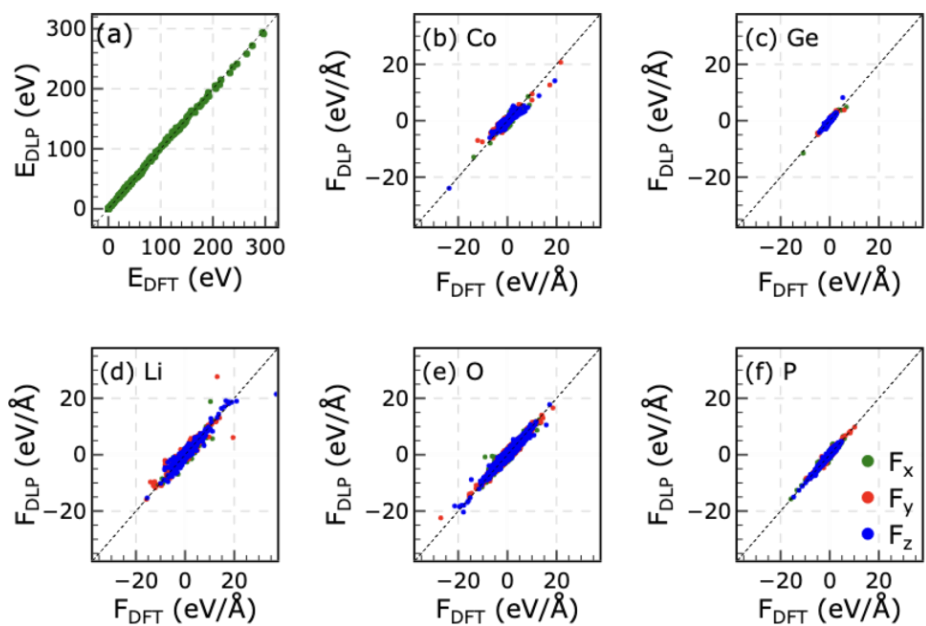

Figure 3: Energy (a) and Forces(b-f) of DFT vs DLP for elements on the LCO|LGPS interface. $F_{DFT}$ and $E_{DFT}$ are the forces and energy, respectively, calculated from DFT-generated data. $F_{DLP}$ and $E_{DLP}$ are the predicted forces and energy, respectively, generated using the DLP.

The plots in Figure 3 predict energy and forces using DLP to validate our model. An ideal correlation number will be 1, given by a perfect line running at a 45-degree angle starting from the origin. Our results show a variation of +/- 0.1% from this line, which is highly accurate.

**Machine Learning Molecular Dynamics (MLMD) simulation**
The MLMD simulations were conducted using the LAMMPS package built into the DeepMD-kit[50,55]. The VNL software was used to build interfaces with more than 6000 atoms for the MLMD simulations. The structures were minimized using the Polak-Ribiére version of the conjugate gradient algorithm[55,56], and the threshold tolerances for energy and force were set to $10^{-18}$ and $10^{-20}$ eV, respectively. Similar to the AIMD simulations, an NVT ensemble controlled by the Nosé–Hoover thermostat was adopted to maintain a temperature of 300 K. An integration time step of $10^{-3}$ ps was used, with a temperature damping parameter of $10^{-2}$. The interface structures were equilibrated for at least 2 nanoseconds(ns).

**Continuum model;**
In order to understand the behavior of the system in larger time and length scale, and provide correlation with experiments, two types of continuum models are developed at two different length scales:

i) A one-dimensional (1D) framework was established for capturing the time-dependent growth of the interdiffusion-induced interphase layer[57,58]. This model focuses on a thinner region close to the interface between the LCO cathode and the LGPS solid electrolyte, where both domains can be approximated as planar regions with variations only along the thickness direction. Mass and momentum balance relations are solved to predict the distribution of species concentration and the evolution of mechanical stress at the interface. The interdiffusion coefficients estimated from the atomistic simulations are used for calculating the distribution of species concentration. A phase field equation is also solved to keep track of the increase in thickness of the interphase region.

ii) A two-dimensional (2D) framework was developed for predicting the impact of the passivating interphase on the overall cell performance[43]. This 2D model captures the variations occurring over a wider domain that encompasses a significant portion of the LCO cathode and the LGPS solid electrolyte microstructure. Charge, mass, and momentum balance relations were solved in the 2D framework for determining the potential, concentration, and stress distribution within the cathode and the solid electrolyte regions[59,60]. Thickness of the passivating interphase layer, as estimated by the 1D continuum model, was used for calculating the charge transfer resistance at the cathode|solid-electrolyte (LCO|LGPS) interface. The thickness of the degraded cathode active material was also assumed to be half of the thickness of the passivating interphase layer.

All the governing equations solved for both the 1D interdiffusion and 2D performance models, and the corresponding list of parameters used for running the simulations, are provided within the Supplementary Information (SI) section.

**Results and Discussion**

In the present study, the chemical and electrochemical stability of the LCO|LGPS interface and its electrochemical performance is investigated using AIMD, MLMD, and continuum modeling techniques. The results are divided into the following sub-sections:
1. Understanding stability and ionic interdiffusion at the LCO|LGPS interface.
2. Understanding the intermixing of ions at the LCO|LNTO interface.
3. Impact of ionic interdiffusion on the final cell performance.

The advantages of incorporating an LNTO-based protective layer on top of the cathode active material (LCO) to minimize the interdiffusion-induced degradation are elaborated.

**LCO|LGPS interface**

The profile of the MLMD simulation, run up to 3.6 ns at ambient temperature, for the (001)LGPS|(010)LCO interface is plotted in Figure 4. The profile shows that potential energy drops by 44 meV from 0 to 2 ns, after which the decrease becomes negligibly small. This indicates that the interdiffusion of Co and O atoms reaches a saturation point or becomes minimal. Similarly, the pressure profile follows the same trend as the potential energy. The temperature fluctuates around the target value throughout the simulation, indicating proper thermal control. The kinetic energy, closely related to temperature,

shows corresponding fluctuations. These profiles collectively suggest that the system approaches equilibrium after about 2 ns, with the initial rapid changes in potential energy and pressure likely corresponding to the primary interdiffusion events at the interface. The stabilization of these properties after 2 ns implies that the interface structure has largely settled into a stable configuration, with only minor atomic rearrangements occurring thereafter.

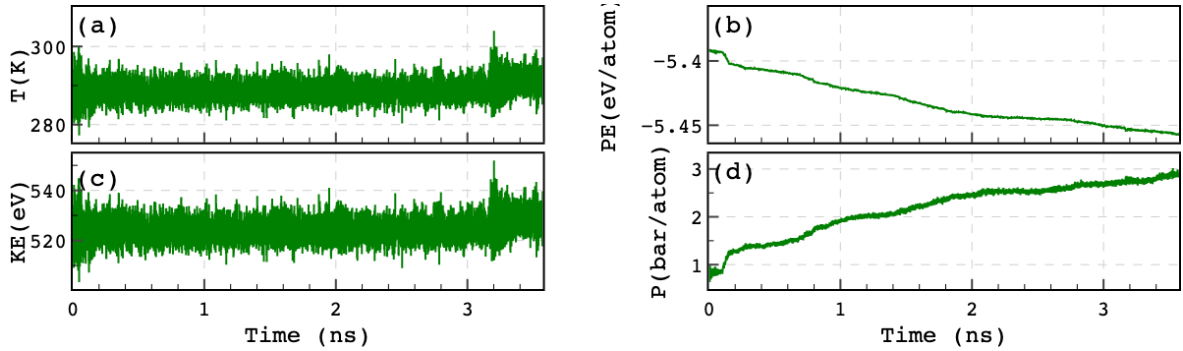

Figure 4 : Large-scale MD simulation of (001)LGPS|(010)LCO up to 3.6ns using Deep Learning Potential.

To design and perform mesoscale level simulations, the diffusion coefficients (D) of all the chemical species associated with the LCO|LGPS interface and the bulk structures (LCO and LGPS) were estimated using equations in S25 and S26 . The calculated MSD and D are shown in Figure 5 and Table 2.

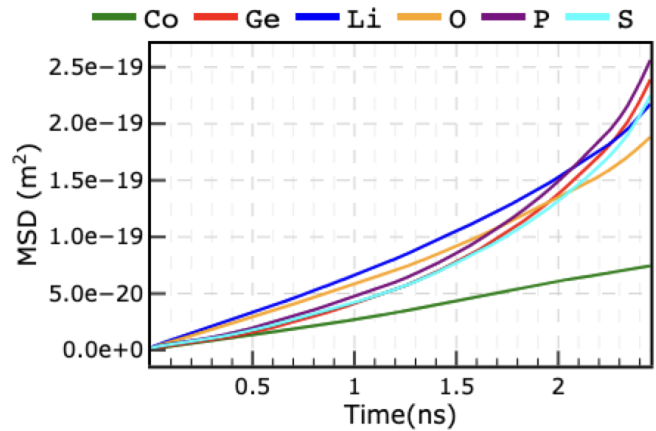

Figure 5: MSD profiles of atomic species associated with (010)LCO|(001)LGPS interface.

Table 2: Diffusion coefficients (D) in $m^2/s$ of elements in LGPS|LCO

| Diffusion Elements | Calculated D (m2/s) | | | | |
|---|---|---|---|---|---|
| | Bulk-LCO | | Bulk-LGPS | | LCO(010)|LGPS(100) |
| | MLMD | Expt[1] | MLMD | Expt[2] | MLMD |
| Li | 5.7E-16 | 1e-16 | 4.5E-10 | 1E-10 | 3E-12 |
| Co | 2.1E-17 | | | | 8E-15 |
| O | 6.2E-16 | | | | 2E-12 |
| Ge | | | 1.06E-12 | | 4E-14 |
| P | | | 8.65E-13 | | 7E-12 |
| S | | | 9.83E-13 | | 6E-13 |

We considered the three LCO surfaces identified in the computational methodology, i.e., 110, 104, and 010, to construct an interface with a 100 LGPS surface. The interdiffusion of ions over a period of 3 ns at the most stable interface, LCO(010) |LGPS (100), is illustrated in Figure 6. The designed LCO (110 and 104) |LGPS interface structures are illustrated in Figure S2.

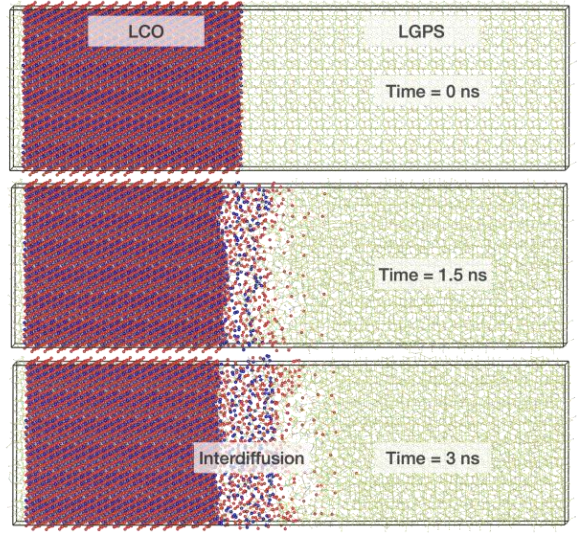

Figure 6: Interdiffusion of Co atom into LGPS. (a) NVT@ 0ns, (b) NVT @ 1.5 ns, (c) NVT@3ns. Red, blue, green, yellow, purple, and gray balls are O, Co, Li, S, P, and Ge atoms, respectively.

An interphase layer is observed to grow steadily with time, and the diffusion of Co to LGPS spans over a limited thickness of the layer. As the simulation progresses, the Co and O concentration increases in the formed layer, which shows a continuous degradation of the LCO cathode. At the same time, there is no significant Co interdiffusion found at (110)LCO|(001)LGPS and (104)LCO|(001)LGPS interfaces

The developed atomistic framework captures the interdiffusion of atoms and ions in the nanosecond timescale and nanometer length scale. However, realistic batteries operate on a timescale of hours to years, and the length scales are in microns, which is consistent with the size of the cathode and solid-electrolyte particles and electrode thicknesses. Accordingly, continuum modeling of the interdiffusion process is conducted at larger length and time scales using the diffusion coefficients extracted from the atomistic simulations. The developed machine learning model described in the Methodology section captures the interdiffusion-induced increase in thickness of the interphase layer, evolution of species concentration within the interphase and adjacent LCO and LGPS domains, and variation in local pressure due to the intermixing of ionic species. The flow of current is not considered during the atomic interdiffusion process. Since we observed a clear Co-interdiffusion from LCO to LGPS at (010)LCO|(001)LGPS interface, we calculated atom-resolved MSD and plotted it in Figure 5. From the MSD values, diffusion coefficients were calculated and compared with the corresponding bulk values in Table 2. At the interface, the trends of atomic diffusion coefficients are O>Li>P>S>Ge>Co. It is understandable that the trend of diffusion coefficients is different at its bulk nature due to the formation of interface.

The calculated parameters from MLMD are imported into the continuum model, and it predicted the evolution of the interphase layer with time, which is shown in Figure 7(a). Initially, a sharp interface exists between LCO and LGPS, as shown by the dashed black line in Figure 7(a). The interphase layer thickness increases with time due to the interdiffusion process, and the model-predicted diffused interphase, after 150 hours, is denoted by the solid black line in Figure 7(a). Location of the phase variable after 3 hours is also depicted in Figure S5(a) by the dashed line. Identification of the LCO, interphase, and LGPS domains is conducted based on the magnitude of the phase variables; anything greater than 0.95 is considered LGPS, a phase variable less than 0.05 is considered LCO, and the intermediate regions indicate the interphase layer. Species interdiffusion leads to the intermixing of ions and atoms within the three domains, where the rate depends on the diffusion coefficients of the individual species. Diffusion-induced concentration profiles of the various species within the three domains after 150 hours are plotted in Figure 7(b). Concentrations of oxygen (O), cobalt (Co), sulfur (S), phosphorus (P), and germanium (Ge) are depicted in Figure 7(b) by the black, blue, red, green, and magenta lines, respectively. Lithium is the common species between the LCO cathode and LGPS solid electrolytes, and their concentrations are estimated by satisfying the charge neutrality. Due to the relatively higher diffusivity of species within the interphase region, no significant concentration gradients are observed within a particular region. However, jump in concentration profiles exist due to the consideration of three different domains within the modeling framework.

Time-dependent increase in the thickness of the interphase layer and local pressure is shown in Figure 7(c) by the solid black line (left vertical axis) and the solid red line (right vertical axis), respectively. It is well understood that interdiffusion of species leads to an increase in interphase layer thickness, which increases rapidly at shorter times (< 10 hours), but slows down eventually at longer times (> 50 hours). The decrease in interphase layer growth rate can be attributed to the increase in local compressive pressure and subsequent retardation of the mass transport processes[61,62]. Compressive stresses usually lead to a decrease in lattice spacing and increases the kinetic energy barrier associated with movement of ions from one location to other, which effectively results in a decrease in diffusion coefficients[63]. The model assumes that interdiffusion-induced intermixing of species leads to a local increase in volume, which eventually generates compressive stresses around the interphase region. Since interdiffusion of ions is the source of stresses within the interphase region, the time dependent increase in local pressure follows a very similar trend as that of the interphase layer thickness. Distribution of pressure along the interphase at 3 hours and 150 hours are demonstrated in Figure S5(b), which clearly shows that the compressive stresses does not change significantly with location, such as, within the LCO, interphase, or LGPS domains. However, the magnitude of stresses increases with time due to enhanced species interdiffusion, which is also reflected in Figure 7(c). A power law fit (see Eq. (5)) to the time dependent growth of the interphase layer thickness is also provided in Figure 7(c) by the cyan dashed line:

**Interphase Layer Thickness** $= \mathbf{0.265 \cdot t^{0.155}}$ ---------------- (5)

A time exponent of 0.155, which is smaller than 0.5, indicates that the growth of the interphase layer can be characterized as a damped diffusion mechanism. The stress-induced decrease in diffusivity is the major reason limiting the movement of ions.

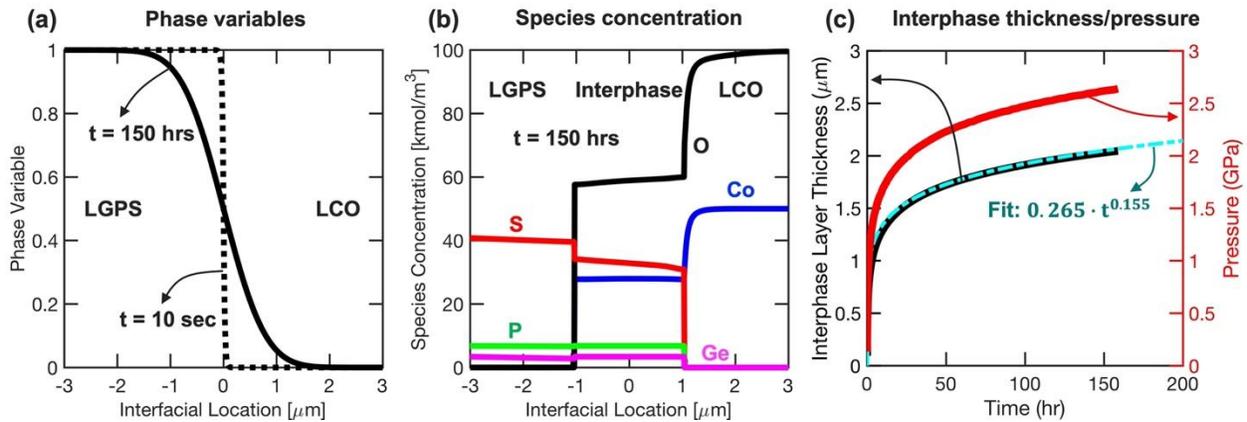

Figure 7. (a) Evolution of the phase variables with time, where LGPS is located on the left, and LCO is on the right side. The sharp interface at 10 sec is shown by the dashed line, whereas the interdiffusion-induced diffused interphase obtained after 150 hours is denoted by the solid black line. (b) The distribution of various species after 150 hours within the LCO, interphase, and LGPS domains. Concentrations of oxygen, cobalt, sulfur, phosphorus, and germanium are shown by the black, blue, red, green, and magenta lines, respectively. (c) Increase in interphase layer thickness and pressure with time is denoted by the solid black (left axis) and solid red (right axis) lines, respectively. A rapid increase in interphase layer thickness is observed initially, which slows down over time due to the compressive pressure-induced decrease in diffusion coefficients. A power law fit to the increase in interphase layer thickness is shown by the cyan line.

**LCO|LNTO interface:**

The MLMD simulations run up to 2ns , for the LCO | LNTO interface, are in agreement with the experimental findings, Co diffusion from LCO is mitigated by LNTO coating. We considered the three LCO surfaces, i.e., 110, 104, and 010, to construct an interface with the 131 LNTO surface. We periodically observed each calculation as it progressed, and the interfaces were stable throughout the simulations. The potential energy profile reached equilibrium within 1 ns for each interface, while the temperature fluctuated

around the target value. The stability of these properties, as illustrated by the potential energy and temperature profiles in Figure S3(a), shows that the system is in a stable configuration. The stability of the LCO cathode was also observed to depend on the orientation of the LCO structure relative to the LNTO. The orientation of LCO, where the Co was aligned in the direction of the interface or at a slight angle, prevented the disintegration of the cathode during simulation. This suggests that the nature of interfacial contact between the two materials plays a role in maintaining the integrity of the structure and mitigating Co diffusion even when a buffer layer is used. Figure 8 illustrates the final trajectories for the investigated LCO|LNTO interfaces at 2ns, and Figure S3(b-d) shows the regular interval observations of each simulation.

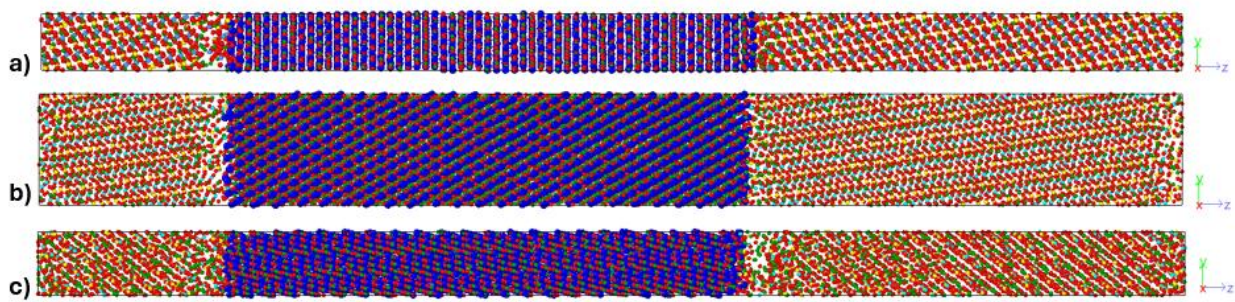

Figure 8: The final trajectory structures at 2ns of LCO|LNTO interface with no interdiffusion of Cobalt(blue) for the LCO surfaces a)104 b) 010 c) 110

The mitigation of Co diffusion is feasible due to the presence of Nb and Ta atoms in the LNTO structure, which form very stable bonds with oxygen, preventing Li substitution by Co. The calculated Li↔Co substitution energy in LNTO was 0.144 eV, whereas the energy of -1.109 eV in LGPS showed a feasible substitution process enabling Co diffusion. Figure S4 illustrates the Li↔Co substitution for both materials. The transition metals (TM) in LNTO have high oxidation states of +5, i.e, $Nb^{5+}$ and $Ta^{5+}$, they form a stable and rigid TM-O bond framework[64,65]. Substituting Li with $Co^{2+/3+}$ creates a charge imbalance, where a charge deficit of +2 to +3 is introduced, as TM redox is not redox-flexible; balancing this charge is energetically unfavorable. Substituting Nb/Ta with $Co^{2+/3+}$ will create an even more significant charge imbalance and is therefore highly unlikely. In contrast, the $Ge^{4+}/P^{5+}$ sulfide framework in LGPS is redox-flexible[65]; therefore, the charge deficit created by substituting Li with $Co^{2+/3+}$ is compensated, making it energetically favorable. The work of adhesion($W_{ad}$) was calculated to evaluate the interfacial binding strength of the interfaces at 300K[66]. The LCO|LNTO interface had $W_{ad}$ of 1.5605 eV /Å$^2$, which was higher than LCO|LGPS (0.7306 eV /Å$^2$). LCO|LNTO, therefore, requires more energy for the interface to be separated into two free surfaces; hence, it is more stable than LCO|LGPS. Due to the inherent stability of the LCO|LNTO interface, no continuum level simulations were run to investigate the extent of ionic interdiffusion between LCO and LNTO. The impact of LNTO interphase layer on the overall electrochemical behavior of the LCO|LGPS systems will be discussed next.

**Analyzing cell performance:**

The continuum modeling technique, described in the Methodology section, estimates the voltage vs. capacity performance curves of the LCO cathode with different electrolytes. The cathode/electrolyte computational mesh, provided in Figure S6(a), was used for this analysis. A slow discharge at 0.1C rate, with different electrolytes, was modeled and compared with experimental results extracted from existing literature [7,27]. During the discharge process, lithium concentration within the cathode increases. The LCO particles go from a lithium-poor configuration at high voltages to a lithium-rich state at lower voltages at the end of discharge. Mass transport limitations lead to the formation of concentration gradients in the LCO cathode. The concentration profile towards the end of discharge, at around 3.5 V, is shown in Figure 9(a). A significant concentration gradient is observed between the surface and the center of the cathode active particle. Formation of such a large concentration gradient, even at a 0.1 C rate, can be attributed to the dramatic decrease in lithium diffusion coefficient in LCO under fully lithiated conditions[67–69]. Heterogeneous grain orientation-induced anisotropic mass transport processes are also considered, which result in nonuniformity in lithium concentration. The LGPS SE acts as a single ion conductor where lithium transport occurs through the migration mechanism, and no concentration gradient evolve within the electrolyte.

Distribution of hydrostatic stresses within the LCO cathodes and LGPS SE due to the evolution of concentration gradients are shown in Figure 9(b). Completely lithiated LCO should provide a stress-free configuration, whereas the extraction of lithium by any amount should lead to the evolution of stresses. The heterogeneity in stress distribution within the bulk of the cathode can be attributed to the anisotropic lattice properties, anisotropic swelling, and randomness in grain orientation within the LCO cathode[70,71]. No grain orientation dependent anisotropy is considered for the LGPS SE. Additionally, the lithium concentration in LGPS does not change at all; accordingly, it does not contribute to the generation of stresses. Even though the hydrostatic stress within the bulk of LCO varies over a range of several gigapascals (-2 GPa to 1.5 GPa), the stress at the interface between LCO and LGPS SE is relatively less, around 50 MPa (see Figure S7(b) and S7(e)). Accordingly, the propensity of interfacial delamination between LCO and LGPS can be minimal. This interfacial mechanical stability with LGPS arises from the mechanical softness of the SE. If LLZO-based stiffer SEs are used, the interfacial stresses can exceed 500 MPa (see Figure S7(e)), resulting in severe delamination between the cathode and SE [72]. Note that the lack of mechanical delamination between LCO and LGPS does not necessarily indicate a completely stable interface between the two, because the interdiffusion-induced degradation is still possible.

Finally, the model predicted voltage vs. capacity discharge curves for the LCO cathodes, while operating with different electrolytes, are demonstrated in Figure 9(c) (solid lines), and compared with the corresponding experimental observations (circular symbols)[7]. The discharge performance of LCO with liquid electrolytes (LE) at 0.1C is shown by the black color. Incorporation of the lithium concentration-dependent decrease in diffusivity is sufficient to obtain a good correlation between the model and experiments[67]. Neither interdiffusion of ions, nor interfacial delamination, are taken into consideration for this comparison of performance with liquid electrolyte. Due to the absence of any detrimental degradation mechanisms, maximum capacity is extracted for the LCO|LE systems.

The experimentally observed discharge curve of LCO while operating with LGPS solid electrolyte is shown by the red circles in Figure 9(c). The dramatic decrease in capacity, even in the first discharge, is attributed to the interdiffusion-induced formation of the passivation layer[7,18]. Loss of cathode active material is also taken into consideration within the interdiffusion-induced interphase region that penetrates into the LCO cathode.

In Figure 7(c), the model predicts rapid interdiffusion of ions and a corresponding increase in the thickness of the passivation layer. Based on the estimated interdiffusion coefficients from atomistic simulations, more than 1 µm thick detrimental interphase layer can evolve within 24 hours of bringing the LCO cathode in contact with the LGPS solid electrolytes. In order to obtain good correlation with experimentally observed capacity, a 1 µm thick passivation layer is assumed within the LCO cathode where the fraction of inactive LCO is assumed to be around 0.6. The fraction of inactive cathode influences the capacity in two ways[36]:

A) the thermodynamic capacity of LCO decreases due to the loss of sites for accommodating lithium
B) the local diffusivity of lithium decreases from the porosity and tortuosity introduced by the inactive LCO, which aggravates the kinetic limitations

Also, the reference exchange current density at the LCO|LGPS interface is decreased by a factor of 0.15 to demonstrate good correlation with the experimentally observed voltage values. Note that due to the inherent softness of the LGPS SE, the interfacial delamination between LCO and LGPS is not considered.

The discharge curve for LNTO coated LCO operating with LGPS is shown in Figure 9(c) by the blue symbols and solid lines[7]. As estimated by atomistic simulations, the presence of LNTO between LCO and LGPS almost eliminates the interdiffusion of ions. Accordingly, the formation of a detrimental passivation layer can be ruled out in the presence of the LNTO interphase. However, it is worth noting that LNTO demonstrates much higher elastic modulus than LGPS[73–76], which can lead to evolution of a larger amount of tensile stresses at the LCO|SE interface (see Figures S7(d) and S7(e) in SI). This enhanced evolution of stresses can lead to interfacial delamination at LCO|LNTO interface. In spite of the lack of interdiffusion-induced formation of the passivation layer, with the incorporation of LNTO, the experimentally observed decrease in discharge voltage and capacity can be attributed to the interfacial delamination process[27]. In the present analysis, good correlation with experiments is obtained by introducing 30% interfacial delamination between LCO and LNTO, which is assumed to be randomly distributed along the LCO|SE interface. Severe delamination between LCO and LNTO

(delamination > 70%) is not considered due to the larger adhesion strength of the LCO|LNTO interface as compared to the LCO|LGPS one. Lower ionic conductivity of LNTO $(1 \times 10^{-4}\,\text{S/m})$ [26] can also contribute to an increase in the charge transfer resistance between LCO and SE, which is incorporated into the model by decreasing the reference exchange current density by a factor of 0.2. Overall, LNTO is capable of successfully eliminating the interdiffusion-induced challenges at the LCO|LGPS interface, but its mechanical stiffness and limited conductivity can still pose issues associated with interfacial delamination during long-term cycling of the cell.

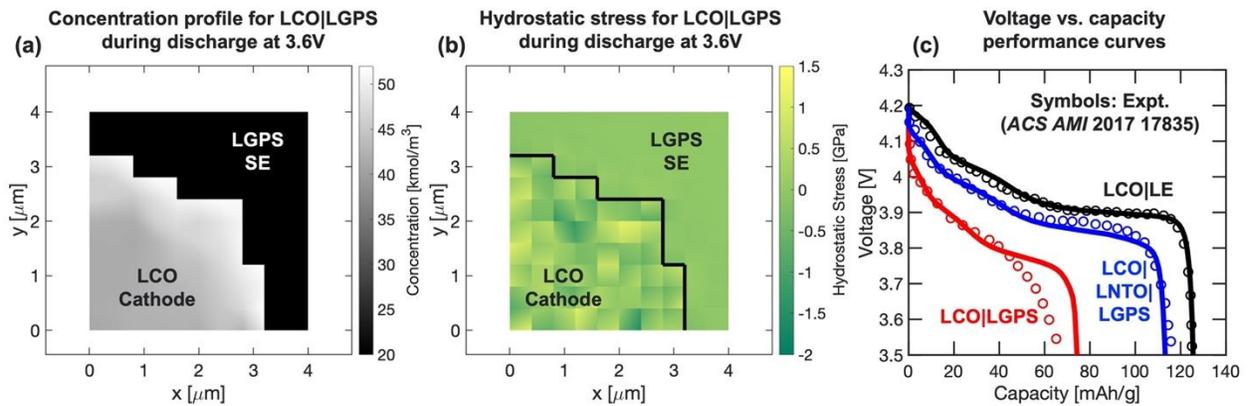

Figure 9. (a) Distribution of lithium concentration within the LCO cathode close to the end of discharge at 0.1C rate. No variation in electrolyte salt concentration occurs within LGPS because it is a single-ion conducting solid electrolyte. (b) Distribution of hydrostatic stress within the LCO cathode and LGPS solid electrolyte as observed close to the end of discharge. The randomness in stress distribution observed within the LCO cathode can be attributed to the randomness in the orientation of the anisotropic grains. (c) Comparison between the model-predicted (solid line) and experimentally observed (circular symbols) voltage vs capacity discharge curves for LCO|liquid-electrolyte (LCO|LE, shown in black), LCO|LGPS (red), and LCO|LNTO|LGPS (blue) systems.

**Conclusion**

In this paper, we utilize the accuracy of ab initio molecular dynamics and the continuum model to investigate the instability observed in high-voltage cathode materials when paired with sulfide-based electrolytes, as reported by experimental methods. We studied the use of an oxide interphase layer, LNTO, to mitigate Co diffusion from the LCO cathode to the LGPS electrolyte. The LCO|LNTO interface showed stability and no Co interdiffusion within the investigated timescale of 2ns. In contrast, LCO|LGPS was

unstable, with ionic interdiffusion observed within 2ns, resulting in the formation of a passive interfacial layer. The thermodynamic calculations provided more clarity on how LNTO and LGPS perform with LCO. The energy required to substitute Li by Co in LGPS was negative, i.e., -1.109 eV, indicating the feasibility of LCO to disintegrate and form other compounds at the interface. In contrast, for LNTO, the energy required was 0.144eV, suggesting a more stable environment. This is also supported by the interfacial binding strengths, which show LCO|LNTO to be more stable with a larger adhesion work than LCO|LGPS. The continuum level analysis revealed that, although LNTO is capable of successfully mitigating the interdiffusion-induced challenges at the LCO|LGPS interface, its mechanical stiffness and limited conductivity can still pose issues associated with interfacial delamination, which may be observed during long-term cycling. This delamination phenomena can limit the oxide interphase layer's ability to act as a protective barrier against ionic interdiffusion. This study explains the experimental observations that, despite their initial success, LNTO cannot completely suppress Co migration up to 300 charge-discharge cycles[27], due to the interfacial degradation through delamination and subsequent ionic interdiffusion.

**Data availability**

Data supporting the work reported in this paper is available within the article and in the supplementary materials. Requests for additional materials should be addressed to the corresponding authors.

**Author contributions**

The DFT, AIMD, and MLMD computational calculations were performed by M.K. Ncube and S.C. Selvaraj under the supervision of A.T. Ngo and L.A. Curtiss. The Continuum model calculations were carried out by P. Barai under the supervision of V. Srinivasan . All authors contributed to writing, reviewing, and editing the manuscript. L. A Curtiss, A.T. Ngo, and V. Srinivasan were responsible for funding acquisition. A.T. Ngo and V. Srinivasan conceptualized this work.

**Acknowledgments**

The work reported in this paper was supported by the Vehicle Technologies Office (VTO), Department of Energy (DOE), USA, through the Battery Materials Research (BMR) program. Argonne National Laboratory is operated for DOE by UChicago Argonne, LLC under the contract number DE-AC02-06CH11357. We also acknowledge the high-performance computing resources provided by the Laboratory Computing Resource Center at the Argonne National Laboratory (ANL).

**Declaration of conflict of interest**

The authors declare no conflict of interest.